\documentclass[conference]{IEEEtran}
\ifCLASSINFOpdf
   \usepackage[pdftex]{graphicx}
\else
   \usepackage[dvips]{graphicx}
\fi
  \usepackage[caption=false,font=footnotesize]{subfig}
\usepackage{url}

\graphicspath{{pdf/}{jpeg/}{png/}{eps/}{figures/}}
\DeclareGraphicsExtensions{.pdf,.jpeg,.png,.eps}
\usepackage{algorithmic}
\usepackage[table]{xcolor}
\usepackage{multirow}
\usepackage{listings}
\usepackage{stfloats}
\usepackage{balance}
\usepackage{soul}
\usepackage[normalem]{ulem}
\usepackage{cancel}

\begin{document}
%
\title{Demo Abstract: A Research Platform for Real-World Evaluation of Routing Schemes in Delay Tolerant Social Networks 
}



\author{\IEEEauthorblockN{Corey E. Baker\textsuperscript{1}, Allen Starke\textsuperscript{2}, Shitong Xing\textsuperscript{1}, Janise McNair\textsuperscript{2}}
\IEEEauthorblockA{
\textsuperscript{1}Department of Electrical and Computer Engineering, University of California, San Diego\\
\textsuperscript{2}Department of Electrical and Computer Engineering, University of Florida\\
Email: cobaker@eng.ucsd.edu, allen1.starke@ufl.edu, sxing@ucsd.edu, mcnair@ece.ufl.edu}}


%


\maketitle

\IEEEpeerreviewmaketitle

\section{Introduction}\label{sec_intro}

Over the past decade, online social networks (OSNs) such as Twitter and Facebook have thrived and experienced rapid growth to over 1 billion users~\cite{faloutsos2010online}. 
%
A major evolution would be to leverage the characteristics of OSNs to evaluate the effectiveness of the many routing schemes developed by the research community in real-world scenarios. 
%
In natural disaster situations, Internet and cellular 
communication infrastructures can be severely disrupted, prohibiting users from 
notifying family, friends, and associates about safety, location, food, water, and 
other resources. 
Disasters typically damage infrastructure, which increases network traffic demands 
on any available undamaged infrastructure, causing congestion and delays.

Opportunistic communication can seamlessly supplement Internet connectivity 
when needed and keep communication channels open even during high-use and 
extreme situations.
DTN routing has the ability to deliver data in an intermittent network, 
but a major challenge for DTN routing is assessing real-world 
performance~\cite{Baker2013}. 
To truly understand the reliability of DTNs and their ability to support social networks, it is imperative that DTN routing schemes are evaluated 
\textit{in vivo} with use-cases that are replicable, comparable, and available 
to a variety of researchers. 

In this demo, we showcase AlleyOop Social, a secure delay tolerant networking research platform that serves as a real-life mobile social networking application for Apple iOS devices. AlleyOop Social allows users to interact, publish messages, and discover others that share common interests in an intermittent network using Bluetooth, peer-to-peer WiFi, and infrastructure WiFi. The research platform serves as an overlay application for the Secure Opportunistic Schemes (SOS) middleware which allows different routing schemes to be easily implemented relieving the burden of security and connection establishment.
AlleyOop Social is named after the basketball play known as an ``alley oop''. 
An ``alley oop'' occurs when one player throws the ball close to the basket, but it 
is not able to reach the final destination. 
While the ball is in flight, a teammate that is closer 
to the basket catches the ball and scores. 
In the same regard, AlleyOop Social enables wireless mobile users to 
communicate over longer distances by sending messages that cannot reach 
the final destination, but are ``caught'' by intermediate mobile devices, 
which continue to catch and pass the messages until they are delivered to 
the final destination.

\begin{figure}
	\centering
	\includegraphics[width=.48\textwidth, height=2.1in]{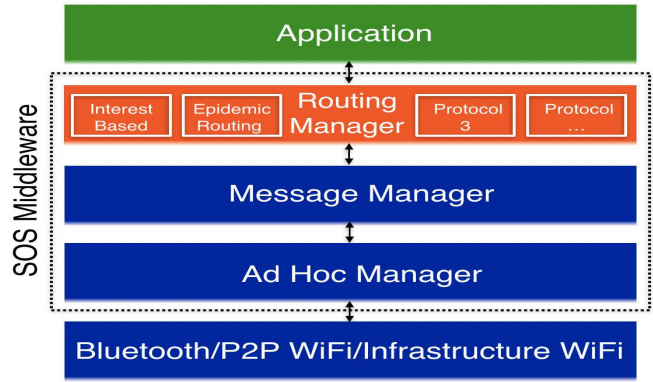}
	\caption{SOS middleware system stack - green represents mobile applications created by developers, orange represents the modular routing layer consisting of multiple opportunistic schemes created by academic researchers, blue represents foundational layers in the middleware that consist of encryption, authentication, peer discovery, connection management, and data dissemination. Managers marked with the color ``blue'' cannot be modified by mobile application developers or academic researchers.}
	\label{fig_alleyoop_sys_stack}
\end{figure}

\section{AlleyOop Social Research Platform}\label{alleyoop_sec_architecture}
AlleyOop Social is designed to feel familiar to users of 
well-known social networking applications such as
Twitter and Facebook.
In addition, AlleyOop Social simultaneously operates as 
an online and offline delay tolerant mobile social network. 
%
Users download the app and create an account, enabling them to develop social 
circles, and then pass and publish messages to each other. 
AlleyOop Social is a research platform allowing users to 
select any DTN routing protocol available in the SOS middleware (discussed in Section~\ref{sec_sos_middleware})
to disseminate messages. 
Researchers can perform controlled studies by asking people 
to select various protocols in AlleyOop Social and post messages.
AlleyOop Social gathers analytics about D2D encounters
with friends and AlleyOop Social users.
Unlike the Haggle project \cite{su2007haggle} which is not fully functional on 
iOS and prefers devices to be rooted on Android, AlleyOop Social gathers 
analytics about encryption/decryption duration, sign/verify duration, the 
app was in: foreground, background, and suspended, along with the size of 
data that was disseminated.
%

\begin{figure*}
\centering
    \subfloat{\includegraphics[width=0.48\textwidth, height=2.49in]{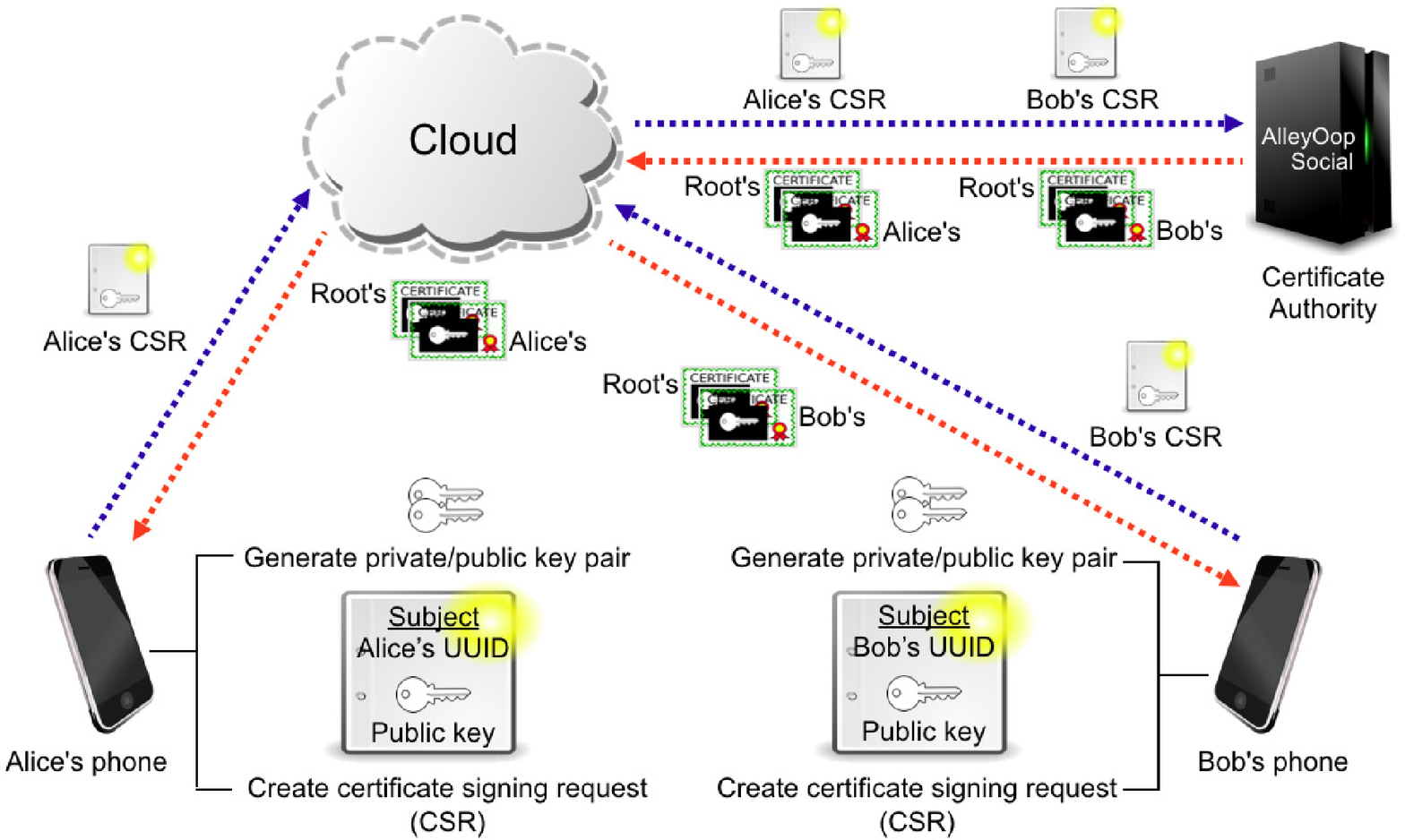}\label{fig_alleyoop_sign_up}} 
    ~
	\subfloat{\includegraphics[width=0.48\textwidth, height=2.49in]{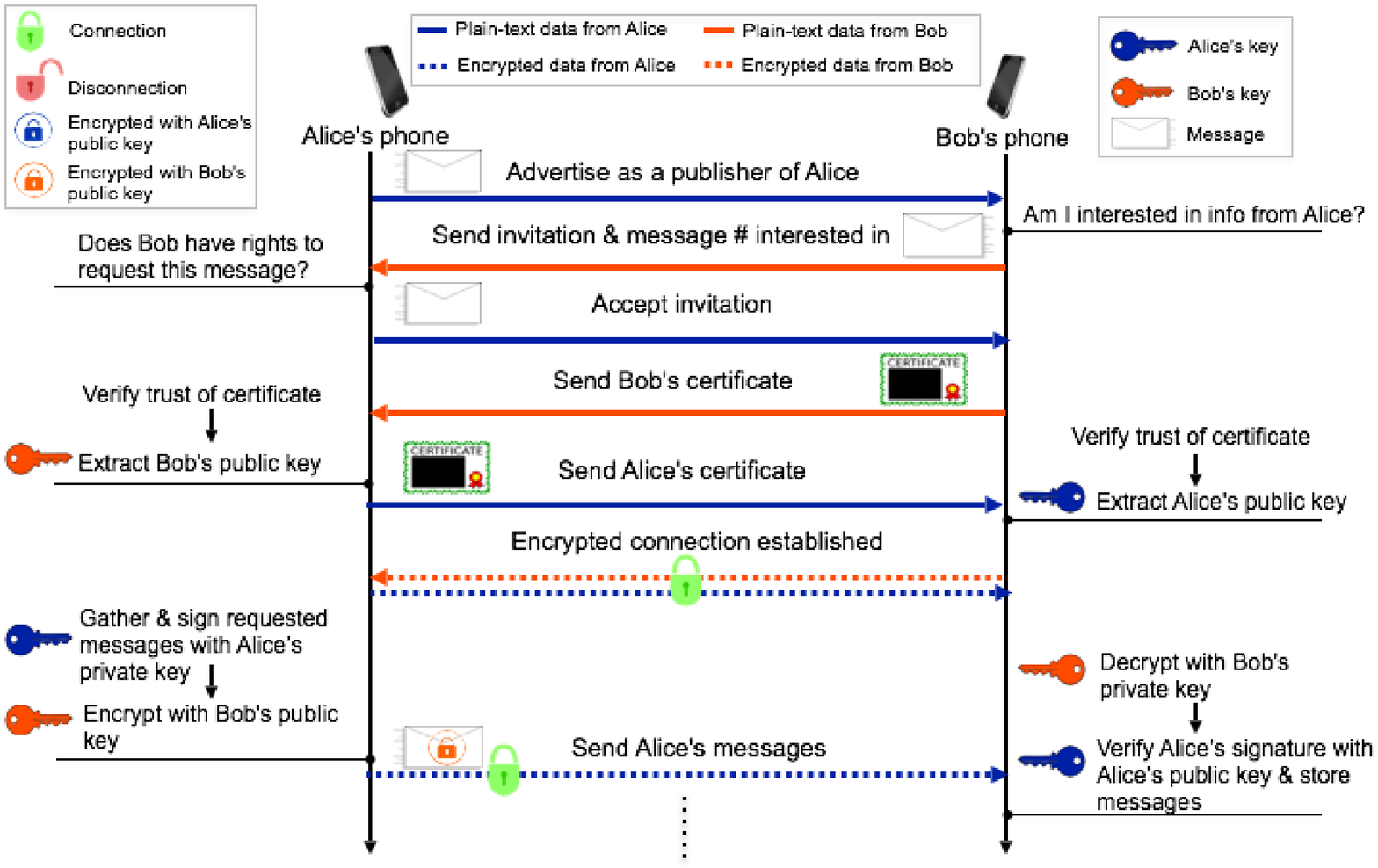}\label{fig_allleyoop_decentralized}} 
   
\caption{(a) One-time infrastructure requirement, occurs during account creation to enable DTN security. (b) Decentralized communication between nodes.} 
\label{fig_alley_security}
\end{figure*}

\section{Secure Opportunistic Schemes (SOS) Middleware}\label{sec_sos_middleware}
The SOS middleware is an underlying framework that turns the AlleyOop Social 
research platform into a delay tolerant mobile social network.
The SOS middleware takes a modular approach to abstract away much of the 
complexity involved in implementing opportunistic routing schemes such as 
device dicovery, establishing D2D connections, and handling device security and 
privacy.
%
%
DTNs are intended to provide an overlay architecture above the existing 
transport layer and ensure reliable routing during intermittency~\cite{Fall2003}.
Building on the knowledge gained from other middleware, SOS hides the complexity of the network stack (session and 
presentation) within the ad hoc manager, message manager, and routing manager,
allowing any mobile application to run at the application layer as depicted in 
Figure~\ref{fig_alleyoop_sys_stack}.
Different from other middlewares such as the Haggle Project~\cite{su2007haggle},
a separate instance of the SOS middleware is intended to run within each mobile application 
as opposed to a daemon which often requires devices to be rooted or jailbroken.
Designing SOS in this manner allows for the middleware to be integrated within individual 
mobile applications in iOS, enabling the overlaying applications to support opportunistic communication 
without jailbreaking devices along with being compliant with App Store regulations.

\section{Privacy and Security}\label{alleyoop_sec_privacy}
In regard to network security there is no ``one-size-fits-all" 
approach~\cite{perlman1999overview}.
Security concerns may become exacerbated in delay tolerant and ad hoc 
applications where nodes are vulnerable to attacks such as eavesdropping, 
denial of service, and compromised devices.
Providing secure communication that prevents an adversary from accessing and/or 
modifying data is a fundamental requirement of any DTN application~\cite{cabaniss2015multi}.
Previous research discusses security in opportunistic applications as a 
proof-of-concept, and makes no claim their implementations are 
secure~\cite{su2007haggle}.
As depicted in Figure~\ref{fig_alley_security}, AlleyOop Social introduces a novel, but simple concept and 
implementation of an initial layer of security for opportunistic communication 
and enables the overlaying mobile application to detect the identity of its 
users, send encrypted information, verify the originating source of the 
information being forwarded, and ensure that data have not been 
modified --- all with minimal dependence on centralized infrastructures.

\section{Demo Setup}\label{alleyoop_sec_demo}
During the demonstration attendees will be able to download AlleyOop 
Social on their respective iOS devices via Apple TestFlight.
Users can follow friends, post new messages, as well as toggle between DTN
routing schemes inside the application.
We will demonstrate both the online and offline modes by disconnecting mobile 
devices from cellular and WiFi networks.
Details of what occurs on every mobile device when a user creates an account and 
disseminates messages is depicted in Figures~\ref{fig_alleyoop_sign_up} and \ref{fig_allleyoop_decentralized} respectively.

\bibliographystyle{IEEEtran}
\bibliography{dtnreferences}
%

\end{document}